\documentclass[journal=jacsat,manuscript=article,layout=twocolumn]{achemso}

\usepackage[version=3]{mhchem} 
\usepackage{amssymb}
\usepackage{amsfonts}
\usepackage{xcolor}
\usepackage{tabularx}

\newcommand{\onlinecite}[1]{[\hspace{-1 ex} \nocite{#1}\citenum{#1}]} 

\author{Lise Morlet-Decarnin}
\affiliation[ENS de Lyon]
{ENSL, CNRS, Laboratoire de Physique, F-69342 Lyon, France}
\author{Thibaut Divoux}
\affiliation[ENS de Lyon]
{ENSL, CNRS, Laboratoire de Physique, F-69342 Lyon, France}
\author{S\'ebastien Manneville}
\email{sebastien.manneville@ens-lyon.fr}
\affiliation[ENS de Lyon]
{ENSL, CNRS, Laboratoire de Physique, F-69342 Lyon, France}

\title[Gelation dynamics in CNC suspensions]
{Critical-like gelation dynamics in cellulose nanocrystal suspensions}


\begin{document}



\begin{abstract} 
We use time-resolved mechanical spectroscopy to offer a detailed picture of the gelation dynamics of cellulose nanocrystal (CNC) suspensions following shear cessation in the presence of salt. CNCs are charged, rodlike colloids that self-assemble into various phases, including physical gels serving as soft precursors for biosourced composites. Here, a series of linear viscoelastic spectra acquired across the sol-gel transition of CNC suspensions are rescaled onto two master curves, that correspond to a viscoelastic liquid state prior to gelation and to a soft solid state after gelation. These two states are separated by a critical gel point, where all rescaling parameters diverge in an asymmetric fashion, yet with exponents that obey hyperscaling relations consistent with previous works on isotropic colloids and polymer gels. Upon varying the salt content, we further show that these critical-like dynamics result in both time-connectivity and time-concentration superposition principles.
\end{abstract}



\section{Main text}
Colloidal nanocrystals are rodlike crystalline clusters of atoms with sizes ranging from tens to a few hundred nanometers \cite{Trache:2020}. These colloids with tunable surface chemistry can be either synthesized or extracted from natural products such as biopolymers. When dispersed into a suspending fluid, colloidal nanocrystals can self-assemble into micro- or meso-structures with outstanding optical and mechanical properties relevant for applications in optics, electronics, sensing, and biomedicine \cite{Parak:2003,boles:2016}. 
Several applications, such as catalysis and optoelectronics, rely on the formation of physical gels, i.e., space-spanning networks of colloidal nanocrystals, which behave mechanically as soft solids. While much is known about the different ways and means to induce gelation in colloidal nanocrystals \cite{Sayevich:2016,Green:2022,berestok:2018}, very few studies have characterized their gelation dynamics and the emergence of solidlike properties upon their self-assembly. Yet, understanding such dynamics is crucial for tailoring the microstructure of nanocrystal gels serving as soft precursors for harder materials.  

Here, we perform a time-resolved mechanical spectroscopy study of the sol-gel transition in cellulose nanocrystals (CNCs). CNCs are biosourced, biodegradable, and biocompatible nanocrystals, which consist of rigid, negatively charged rodlike particles of typical length 100--500~nm, and diameter 5--20~nm \cite{Habibi:2010,Lahiji:2010,Dufresne:2013,Lagerwall:2014,Klemm:2018,Trache:2020,Li:2021,Yucel:2021}.
Aqueous dispersions of CNCs display a rich phase diagram including liquid crystalline phases, gels, and glasses \cite{Xu:2018,Xu:2020}. In practice, in dilute CNC suspensions, gelation is induced by adding salt, hence screening the electrostatic repulsion between the CNCs that aggregate via hydrogen bonds and van der Waals interactions \cite{Cherhal:2015,Peddireddy:2016,Moud:2020}. In this paper, using time-resolved mechanical spectroscopy, we aim to provide a detailed dynamical picture of the sol-gel transition of CNC dispersions following flow cessation in the presence of salt. Following previous works on \textit{time-connectivity} superposition, also referred to as time-cure superposition\cite{Winter:1986,Chambon:1987,Martin:1988,Adolf:1990}, viscoelastic spectra acquired across the sol-gel transition are rescaled onto two remarkable master curves that extend on each side of a critical gel point. These master curves are compactly described by two fractional mechanical models, which  correspond respectively to a viscoelastic liquid state and to a soft solid state, and which share a common element capturing the power-law rheology of the CNC dispersion at the gel point. Moreover, varying the salt content within the CNC dispersion reveals that the pre-gel viscoelastic liquid and the post-gel viscoelastic solid can also be rescaled onto two universal master curves, following a \textit{time-concentration} superposition principle. Finally, we discuss the exponents that characterize the present critical-like gelation dynamics in view of the literature on other chemical and physical gels.

\begin{figure*}[!t]
    \centering
    \includegraphics[width=1\textwidth]{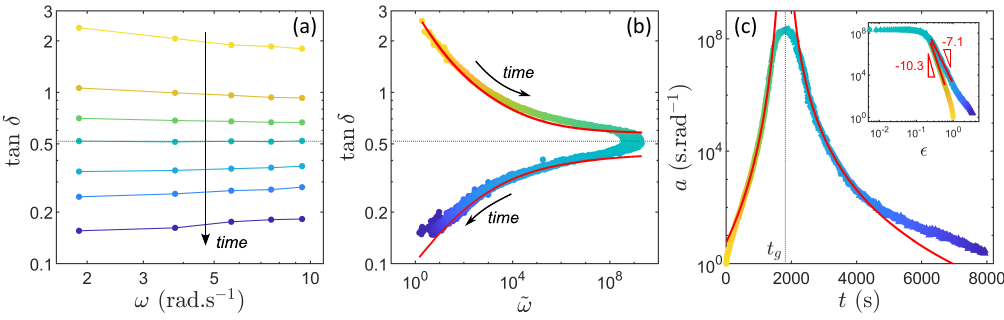}
    \caption{Time-resolved mechanical spectroscopy of the sol-gel transition in a CNC suspension. (a)~Dependence of the loss factor $\tan \delta$ on the frequency $\omega$ at different points in time ($t=$10 to 8000~s, from yellow to dark blue) across the sol-gel transition of a 2\%~wt CNC suspension containing 15~mM of NaCl. The gel point is highlighted by the horizontal dashed line. (b)~Master curve for the loss factor $\tan \delta$ vs reduced frequency $\tilde \omega = a(t)\omega$. The red curves show the best fits of the data respectively by a fractional Maxwell model for $t<t_g$ (upper curve) and by a fractional Kelvin-Voigt model for $t>t_g$ (lower curve). (c)~Time dependence of the shift factor $a$. The initial value is arbitrarily taken as $a(0)=1$~s.rad$^{-1}$. The vertical dashed line highlights the gelation time $t_g=1820$~s. Inset: $a$ vs $\varepsilon=|t-t_g|/t_g$ in logarithmic scales. The red curves in both the main graph and the inset show the best power-law fits of the data, $a\sim\varepsilon^{-y}$, with exponents $y_l=10.3$ for $t<t_g$ and $y_g=7.1$ for $t>t_g$.}
    \label{fig:tan_delta_15mM}
\end{figure*}

CNC gels are prepared using a commercial 6.4\%~wt aqueous suspension of CNCs extracted from wood (CelluForce). The suspension is diluted with salt water to obtain samples containing 2\%~wt of CNCs and salt (NaCl, Merck) in  amounts ranging from 12~mM to 22~mM. The sample is homogenized under high shear using mechanical stirring at 2070 rpm (IKA RW 20 Digital mixer equipped with an R1402 blade dissolver) before, during, and after salt addition for about 5~min at each step. Finally, the sample is stored in the refrigerator for 24~h before being used. At a concentration of 2\%~wt, CNCs are expected to overlap, and in the above range of salt concentrations, CNCs form colloidal gels (see also Supporting Information for more experimental details) \cite{Cherhal:2015,Xu:2018,Xu:2020}. The mechanical properties of the present CNC suspensions are probed during gelation using a stress-controlled rheometer (AR-G2, TA Instruments) equipped with a smooth cylindrical Taylor-Couette geometry (height 58~mm, inner rotating cylinder of radius 24~mm, outer fixed cylinder of radius 25~mm, and gap $e=1$~mm). The cell is closed by a homemade lid, and the temperature is controlled to $T=23\pm 0.1$~$^{\circ}$C, thanks to a water circulation around the cell. This setup allowed us to study the same sample over several hours without any artifact due to evaporation.

After being loaded in the shear cell, each sample is first fully fluidized under a high shear rate $\dot \gamma = 1000$~s$^{-1}$ during 60~s, before being quenched by setting $\dot \gamma=0$, which defines the time origin $t=0$~s. Upon the cessation of shear, the initially liquid CNC suspension slowly reassembles into a physical gel. This sol-gel transition is monitored over up to $5\times 10^4$~s thanks to time-resolved mechanical spectroscopy \cite{Mours:1994}. In practice, cycles of small-amplitude oscillatory stress measurements are performed over five discrete frequencies $\omega$ ranging between 0.3 and 1.5~Hz. This yields one viscoelastic spectrum, defined by the elastic modulus $G'$ and the viscous modulus $G''$ as a function of $\omega$, at every $\delta t_{\rm exp}=5$~s. These frequencies are purposely chosen such that the sample properties do not evolve significantly over this time scale, i.e., $(\delta t_{\rm exp}/G')\,(\partial G'/\partial t)\ll 1$ \cite{Winter:1988,Mours:1994}. 

Figure~\ref{fig:tan_delta_15mM} illustrates the gelation dynamics of a 2\%~wt CNC suspension containing 15~mM of salt. The frequency dependence of the loss factor $\tan \delta =G''/G'$ is reported in Fig.~\ref{fig:tan_delta_15mM}(a) at various points in time across the sol-gel transition. This first allows us to identify the ``true'' gel point \cite{Winter:1997} defined as the time $t_g=1820$~s when $\tan \delta$ is frequency independent [see gray dashed line at $\tan \delta = 0.52$ in Fig.~\ref{fig:tan_delta_15mM}(a)]. Second, $\tan \delta (\omega)$ shows two opposite trends on each side of the gel point: it decreases with $\omega$ for $t<t_g$, while it increases with $\omega$ for $t>t_g$.
We also note that the slope of $\tan \delta$ vs $\omega$ continuously goes from negative to positive across the gel point. This prompts us to apply a time-dependent multiplicative factor $a(t)$ to the frequency $\omega$, in order to collapse the $\tan \delta$ data measured at different points in time onto a single curve \cite{Lennon:2023}, thus revealing the systematic dynamics of the gelation process. As shown in Fig.~\ref{fig:tan_delta_15mM}(b), this rescaling leads to two master curves, one on each side of the gel point, composed of more than 1600 spectra spanning nine orders of magnitude in rescaled dimensionless frequency $\tilde \omega=a(t)\omega$, and describing the entire gelation process. This result points to a time-connectivity superposition principle, as previously identified in polymer gels and colloidal gels made of spherical, rodlike, or fiber-like particles  \cite{Adolf:1990,Larsen:2008,Larsen:2008b,Chen:2010,Hong:2018,Keshavarz:2021,Bantawa:2023}, and highlights the self-similar evolution of the sample viscoelastic properties on each side of the gel point. 
Quite remarkably, the time-dependent shift factor $a(t)$ displays a power-law divergence in the vicinity of the gel point, with a critical exponent $y_l=10.3$ for $t<t_g$ and $y_g=7.1$ for $t>t_g$ [see Fig.~\ref{fig:tan_delta_15mM}(c)].

\begin{figure}[!t]
    \centering
    \includegraphics{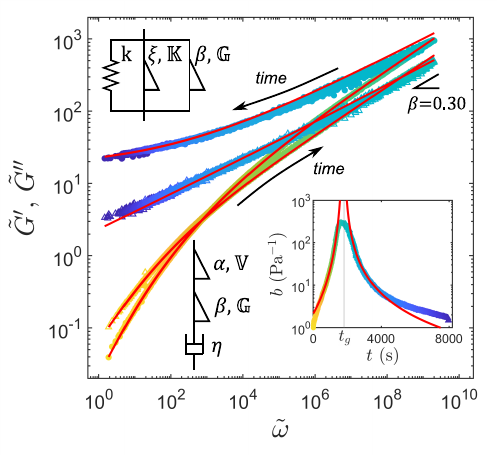}
    \caption{Master curves obtained by rescaling the elastic modulus $G'$ ($\bullet$) and the viscous modulus $G''$ ($\triangle$) by the same multiplicative factor $b(t)$, i.e., $\tilde{G}'=b(t)G'$ and $\tilde{G}''=b(t)G''$. The horizontal axis is the rescaled frequency $\tilde \omega=a(t)\omega$ as defined in Fig.~\ref{fig:tan_delta_15mM}. The red lines show the best fits of the data by a fractional Maxwell model for $t<t_g$ and by a fractional Kelvin-Voigt model for $t>t_g$ (see sketches of the mechanical models as insets). Inset: time dependence of the shift factor $b$. The initial value is arbitrarily taken as $b(0)=1$~Pa$^{-1}$. The red curves show the best power-law fits of the data, $b\sim\varepsilon^{-z}$ where $\varepsilon=|t-t_g|/t_g$, with exponents $z_l=3.0$ for $t<t_g$ and $z_g=1.9$ for $t>t_g$. The vertical dashed line highlights the gelation time $t_g=$1820~s. Same experiment as in Fig.~\ref{fig:tan_delta_15mM}.}
    \label{fig:G_15mM}
\end{figure}

As shown in Fig.~\ref{fig:G_15mM}, we further construct a master curve for both viscoelastic moduli $G'$ and $G''$ vs $\tilde \omega$ by shifting each instantaneous viscoelastic spectrum vertically thanks to a multiplicative coefficient $b(t)$.
These master curves span over five orders of magnitude in rescaled moduli $\tilde{G'}=b(t)G'$ and $\tilde{G''}=b(t)G''$, 
which confirms that time-connectivity superposition applies. Here again, the shift factor $b(t)$ follows critical-like dynamics around $t_g$, yet with exponents $z_l=3.0$ for $t<t_g$ and $z_g=1.9$ for $t>t_g$ that are about three times smaller than those found for $a(t)$ [see inset in Fig.~\ref{fig:G_15mM}].
Moreover, the nine orders of magnitude covered in rescaled frequencies indicate a very wide range of distinct relaxation processes in the sample microstructure. Such a broad relaxation spectrum is often compactly described by fractional models \cite{Jaishankar:2013,Bonfanti:2020,Keshavarz:2021}, which introduce ``spring-pots'' as key rheological elements. A spring-pot is defined by a constitutive equation that relates the stress $\sigma$ and the strain $\gamma$ through a fractional derivative \cite{Jaishankar:2013}, $\sigma=\mathbb{G} \,d^\beta\gamma/dt^\beta$, where $\mathbb{G}$ is a ``quasi-property'' with dimension Pa.s$^{\beta}$, and $\beta\in [0,1]$ is the order of the derivative. In the limit $\beta \rightarrow 0$ (resp. $\beta \rightarrow 1$), the spring-pot corresponds to a purely elastic (resp. viscous) response. For $0<\beta<1$, it displays a power-law viscoelastic spectrum $G'\sim G''\sim \omega^\beta$, or equivalently a power-law relaxation modulus $G(t)\sim t^{-\beta}$, and a frequency-independent phase angle $\delta = \beta\pi/2$. 
Such power-law rheology is common to all ``critical gels,'' that form self-similar percolated networks at the gel point \cite{Winter:1986,Zaccone:2014}, and for which the exponent $\beta$ is referred to as the critical relaxation exponent \cite{Winter:1987,Suman:2020}.
Therefore, the fractional approach is ideally suited to characterize the mechanical response of colloidal gels, from their critical gel point and beyond \cite{Geri:2018,Bouzid:2018, Keshavarz:2021}.

Here, we fit the master curves on each side of the gel point by two five-parameter fractional models, respectively a fractional Maxwell model for $t<t_g$, which captures the liquid-like viscoelasticity of the CNC suspension prior to the gel point, and a fractional Kelvin-Voigt model for $t>t_g$, which captures the solid-like viscoelastic behavior past the gel point (see, respectively, lower and upper sketches in Fig.~\ref{fig:G_15mM}). The reader is referred to the Supporting Information for full mathematical details on the models. A crucial result is that both models share a common spring-pot element $(\mathbb{G}, \beta)$, which is alone responsible for capturing the gel point, here with a critical relaxation exponent $\beta=0.30$ that corresponds to a value of $\tan(\beta\pi/2)= 0.51$, which is fully consistent with the value of $\tan \delta$ observed at the gel point in Fig.~\ref{fig:tan_delta_15mM}(a). The fits by the two models are shown with red lines in Fig.~\ref{fig:G_15mM} for the rescaled viscoelastic moduli, and in Fig.~\ref{fig:tan_delta_15mM}(b) for the corresponding loss factor. The agreement between theory and experiment is excellent, which provides strong support for interpreting the gelation dynamics in terms of two consecutive fractional mechanical behaviors separated by a critical gel point.

\begin{figure*}[!t]
    \centering
    \includegraphics[width=1\textwidth]{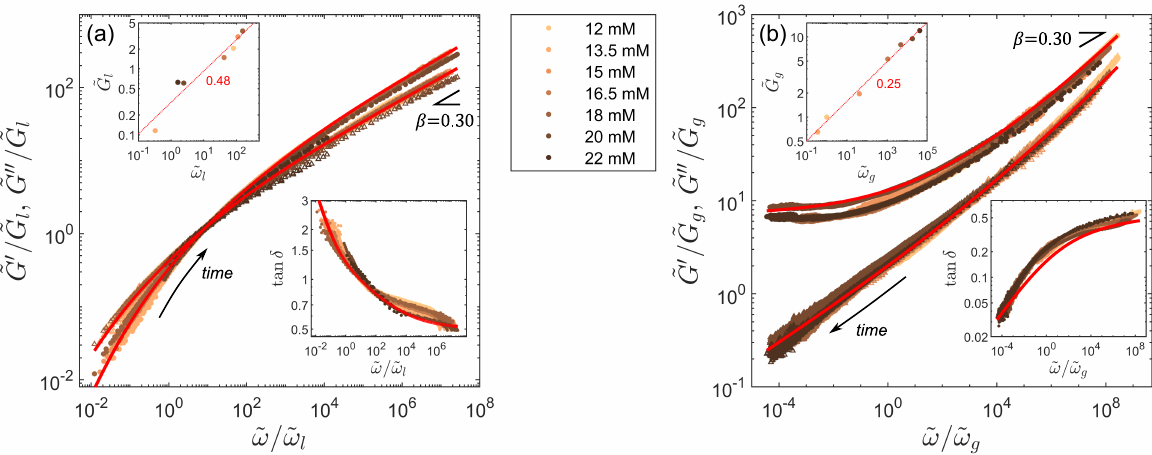}
    \caption{Rescaled master curves for the viscoelastic spectra measured (a) before and (b) after the gel point on CNC suspensions for salt concentrations ranging between 12~mM and 22~mM. Lower insets: $\tan \delta = \tilde G''/\tilde G'= G''/G'$ vs~$\tilde \omega / \tilde \omega_{i}$. Upper insets: shift factors $\tilde G_{i}$ vs $\tilde \omega_{i}$ for the various salt concentrations, together with their best power-law fits (red dotted lines). The indices $i=l$ and $i=g$  respectively denote the pre-gel liquid state in (a) and the post-gel solid state in (b). The red curves show the best fits of the data with a fractional Maxwell model for $t<t_g$ in (a) and with a fractional Kelvin-Voigt model for $t>t_g$ in (b). The data for a salt content of 12~mM were chosen as a reference to construct the master curves. Color code for the concentration as indicated in the legend. Elastic and viscous moduli are shown with $\bullet$ and $\triangle$ respectively.}
    \label{fig:general_master_curve}
\end{figure*}

In order to explore the impact of the salt content on the master curves reported in Fig.~\ref{fig:G_15mM}, the above analysis was repeated for CNC dispersions with salt concentrations ranging between 12~mM and 22~mM. In all cases, we can unambiguously identify a critical gel point associated with a gelation time $t_g$. Upon increasing the salt concentration, the gelation dramatically accelerates due to the stronger screening of electrostatic interactions [see Supporting Fig.~S1(a)], as already reported not only in CNC suspensions \cite{Peddireddy:2016,Morlet-Decarnin:2022}, but also for other types of colloids \cite{Reerink:1954,Linden:2015}. Yet, for all salt contents, a master curve similar to that of Fig.~\ref{fig:G_15mM} can be built, for which the above fractional approach provides very good fits (see Supporting Fig.~S2). Strikingly, at the gel point, the power-law dependence of the viscoelastic moduli with frequency is independent of the salt content, with an exponent $\beta=0.30\pm0.03$ [see Supporting Fig.~S1(b)]. This demonstrates the robustness of both the time-connectivity superposition principle and the microstructure of the percolated network formed at the gel point to a change in the strength of electrostatic interactions between the CNCs. Indeed, the exponent $\beta$ can be related to the fractal dimension $d_f$ of the stress-bearing network at the gel point \cite{Muthukumar:1989, Keshavarz:2021}. Using the relation proposed in Ref.~\onlinecite{Muthukumar:1989} for screened interactions, we obtain $d_f=2.2$ for $\beta=0.30$, which is compatible with independent neutron and light scattering measurements that yield $1.6\lesssim d_f \lesssim 2.1$ \cite{Cherhal:2015}.  

Furthermore, we note that beside $\beta$, the fractional derivative orders $\alpha$ and $\xi$ of the two other spring-pots involved in our fractional models, which respectively control the low-frequency viscoelastic behavior of the viscoelastic liquid for $t<t_g$ and of the
soft solid for $t>t_g$, do not significantly depend on the salt content either (see Supporting Table~S2). This suggests again rescaling all master curves, first by collapsing all loss factors $\tan \delta (\tilde \omega)$ thanks to a simple rescaling of $\tilde \omega$ into $\tilde \omega/\tilde \omega_i$ (see lower insets in Fig.~\ref{fig:general_master_curve}), and then by normalizing both rescaled viscoelastic moduli $\tilde G'$ and $\tilde G''$ with a factor $\tilde G_i$, where $i=l$ ($i=g$ resp.) for the liquid (gel resp.) state. As shown in Fig.~\ref{fig:general_master_curve}, the result is two remarkable universal master curves for the dynamics both before and after the gel point, spanning over twelve orders of magnitude in rescaled frequency, and four decades in viscoelastic moduli. These master curves are consistently nicely fitted by a fractional Maxwell model for $t<t_g$ and by a fractional Kelvin-Voigt model for $t>t_g$, with  $\beta=0.30$, $\alpha=0.64$, and $\xi=0.19$. The shift factors $\tilde G'_i$ and $\tilde\omega_i$ appear to be linked by two different, non-trivial power laws with exponents of roughly $1/2$ in the pre-gel state and $1/4$ in the post-gel state (see upper insets in Fig.~\ref{fig:general_master_curve}). 

Finally, the superposition principles revealed in the present experiments are derived from the critical-like dynamics of the gelation process around the gel point that are characterized by four critical exponents, which values are independent of the salt content [see Supporting Fig.~S1(c)]: $y_l=9.4\pm 0.5$, $y_g=7.4\pm 0.5$, $z_l=2.8\pm 0.2$, and $z_g=2.1\pm 0.2$ when averaged over all concentrations in NaCl. These exponents are linked to those introduced classically in the literature on chemical and physical gels based on percolation theory \cite{deGennes:1976,Adam:1981,Stauffer:1982,Winter:1987,Adolf:1990,Axelos:1990,Hodgson:1990,Winter:1997,Larsen:2008,Rouwhorst:2020,Suman:2020}. In particular, the exponents $y_l$ and $y_g$ associated with the frequency shift factor $a(t)$ in Fig.~\ref{fig:tan_delta_15mM} correspond to the divergence of the longest relaxation time in the system, respectively before and after the gel point. Here, while most previous works have postulated a symmetric divergence, i.e., $y_l=y_g$, we find that the pre-gel exponent is systematically larger than its post-gel counterpart by about 20~\%, out of the range of experimental uncertainty. A similar difference is found between the exponents $z_l$ and $z_g$ derived in Fig.~\ref{fig:G_15mM} from the shift factor $b(t)$ for viscoelastic moduli. Whether such asymmetry in the critical behavior close to gelation is specific to CNCs or general to rodlike colloids is an open issue. Moreover, $z_l$ and $z_g$ relate to the exponents associated with the pre-gel divergence of the zero-shear viscosity, $\eta_0\sim\varepsilon^{-s}$, and with the post-gel growth of the zero-frequency elastic modulus, $G_e\sim\varepsilon^{z}$, through $z_g=z$ and $z_l=y_l-s$. Based on similarity arguments, the exponents $s$ and $z$ were shown to be linked to the rheological exponent $\beta$ at the gel point through two hyperscaling relations, \cite{Stauffer:1982,Winter:1987} $y_l=s/(1-\beta)$ and $y_g=z/\beta$, which simply rewrite $\beta=z_l/y_l=z_g/y_g$ in the present notations. Using the above average values of the various exponents, we get $z_l/y_l\simeq 0.29$ and $z_g/y_g\simeq 0.28$, very close to $\beta\simeq 0.30$, and thus consistent with the predicted hyperscaling and with previous experimental results for micro-rheology on colloidal rods \cite{He:2021}. 


To conclude, our results demonstrate that the gelation of CNC suspensions after shear cessation involves critical-like dynamics, where the salt content drives only the aggregation kinetics. While similar criticality has already been reported many times in chemical and physical gels, our experiments address the case of rodlike colloids for the first time thanks to a systematic rescaling approach, which does not require estimation of a zero-shear viscosity and a zero-frequency modulus by extrapolating viscoelastic spectra at low frequencies. The present study also highlights several peculiarities, including asymmetry in the pre-gel and post-gel exponents and a surprisingly robust time-concentration superposition principle, which call for a microscopic description of the evolution of the colloidal nanorod network across the sol-gel transition, e.g., through time-resolved small-angle scattering. Such complementary microstructural information will help identify a possible universality class for gels made of colloidal nanorods based on the critical exponents and on the fractal features of the space-spanning network. 

\section{Supporting Information}
Supporting text, figures, and tables providing additional details on the materials and methods, on fractional mechanical models, on the data analysis, and on the fitting procedures (PDF). 

\section{Acknowledgments}

The authors thank I.~Capron, B.~Jean, and F.~Pignon for fruitful discussions. This work was funded by the Institut Universitaire de France (IUF). L.~M.-D. also acknowledges financial support from \'Ecole Normale Sup{\'e}rieure de Lyon.


\providecommand*\mcitethebibliography{\thebibliography}
\csname @ifundefined\endcsname{endmcitethebibliography}
  {\let\endmcitethebibliography\endthebibliography}{}


\clearpage
\newpage
\setcounter{equation}{0}
\setcounter{figure}{0}
\global\def\thefigure{S\arabic{figure}}
\setcounter{table}{0}
\global\def\thetable{S\arabic{table}}

\begin{center}
    {\Large\bf{\sc Supporting Information}}
\end{center}

\subsection*{\large Materials and methods}

\paragraph*{CNC samples.-}The rodlike colloids used in this work have an average length $L\simeq 120$~nm and an average diameter $D\simeq 10$~nm, with standard deviations of 40~nm and 2~nm, respectively, as determined from the analysis of transmission electron microscopy images. In the commercial 6.4~wt~\% aqueous suspension from CelluForce, CNCs may be found both as individual nanorods of diameter 3--4~nm and as bundles composed of a couple of nanorods arranged parallel to each other.

Moreover, one may roughly estimate the concentration over which CNCs should overlap by considering an effective volume fraction $\varphi_\mathrm{eff}$ based on the volume fraction of spheres of diameter $L$, namely $\varphi_\mathrm{eff}=\pi L^3 n /6$, where $n$ the CNC number density. This effective volume fraction is linked to the actual CNC volume fraction, $\varphi=\pi L D^2 n/ 4$, through $\varphi=(3D^2/2L^2)\varphi_\mathrm{eff}$. Assuming that CNCs overlap when $\varphi_\mathrm{eff}$ is above some critical jamming volume fraction $\varphi^\star$, and neglecting polydispersity, one gets the following expression for the overlap weight concentration $\varphi_\mathrm{w}^\star$:
\begin{equation}
    \varphi_\mathrm{w}^\star=\frac{\rho_\mathrm{CNC}\ D^2}{\frac{2L^2}{3\varphi^\star}\,\rho_\mathrm{w} +\ (\rho_\mathrm{CNC}-\rho_\mathrm{w})D^2}\,,
\end{equation}
where $\rho_\mathrm{w}=1$~g.cm$^{-3}$ is the water density and $\rho_\mathrm{CNC}=1.5$~g.cm$^{-3}$ the CNC density. Taking $\varphi^\star=0.64$ for random close packing, one gets $\varphi_\mathrm{w}^\star=1$~wt~\%. Therefore, the present system with $\varphi_\mathrm{w}=2$~wt~\% lies above the overlap concentration. Note, however, that the above estimate for $\varphi_\mathrm{w}^\star$ is most likely very crude, because CNCs are highly charged colloids, and the influence of the electrostatic double layers should be taken into account when computing the excluded volume \cite{Xu:2020}.

\paragraph*{Rheological protocol.-}Time-resolved viscoelastic measurements are performed under an imposed oscillatory torque with a ``multiwave method,'' originally developed by Mours and Winter \cite{Mours:1994}, in which the input torque is constituted of the sum of five sinusoids with frequencies  $f=0.3$, 0.6, 0.9, 1.2, and 1.5~Hz, and amplitudes 5, 4, 3, 2, and $1~\mu$N.m, respectively. This allows us to acquire viscoelastic spectra every $\delta t_{\rm exp}=5$~s. For the gel constituted of 2~wt~\% CNC and 15~mM NaCl, the corresponding maximum stress is 22.3~mPa and the maximum strain is 4.5~\%. Such a strain amplitude lies well within the linear viscoelastic domain, which extends up to strain amplitudes of 29~\% as determined from strain sweep experiments. Table~\ref{tab:param} gathers similar parameters for the various gels investigated in the main text.

\begin{table*}[h]
    \centering
\begin{tabularx}{\textwidth}{| >{\centering\arraybackslash}X | >{\centering\arraybackslash}X | >{\centering\arraybackslash}X | >{\centering\arraybackslash}X | }
    \hline
    $[\text{NaCl}]$ (mM) & Maximum stress (mPa) & Maximum strain (\%) & $\gamma_{NL}$ (\%) \rule[-7pt]{0pt}{20pt}\\
    \hline
    12 & 22.1 & 4.6 & 28 \\
    13.5 & 21.8 & 4.6 & 36 \\
    15 & 22.3 & 4.5 & 29 \\
    16.5 & 22.6 & 4.5 & 20 \\
    18 & 22.7 & 4.4 & 14 \\
    20 & 22.8 & 4.1 & 9.0 \\
    22 & 22.8 & 3.7 & 9.0 \\
    \hline
\end{tabularx}
    \caption{Maximum stress and strain values reached upon imposing a multi-frequency oscillatory torque to the various 2~wt~\% CNC gels involved in the present study. The salt concentration is indicated in the first column. The last column gives the strain value $\gamma_{NL}$ that characterizes the extent of the linear viscoelastic regime. $\gamma_{NL}$ is inferred from a strain sweep test, and defined as the strain amplitude above which the elastic modulus has dropped by more than 10~\% compared to its value at rest.}
    \label{tab:param}
\end{table*}

\clearpage
\newpage

\subsection*{\large Fractional rheological models}

As discussed in the main text, we take advantage of fractional mechanical approaches to model the rescaled viscoelastic moduli $\tilde{G}' (\tilde{\omega})$ and $\tilde{G}''(\tilde{\omega})$ of CNC suspensions obtained across the sol-gel transition following flow cessation. Building upon a first spring-pot element ($\mathbb{G}$, $\beta$) that accounts for the sample power-law response at the critical gel point, the liquid-like response of the CNC suspension for $t<t_g$ is described by a fractional Maxwell model (see lower sketch in Fig.~2 in the main text) composed of two additional elements in series with the spring-pot ($\mathbb{G}$, $\beta$), namely a purely viscous dashpot (defined by its viscosity $\eta$) and a second spring-pot ($\mathbb{V}$, $\alpha$) with $0<\beta<\alpha<1$. The corresponding complex viscoelastic modulus reads \cite{Jaishankar:2013}:
\begin{eqnarray}
    \tilde{G}^*(\tilde{\omega})&=&\tilde{G}'(\tilde{\omega})+i\tilde{G}''(\tilde{\omega})\\
    &=&\frac{\mathbb{V} \mathbb{G}\eta (i \tilde{\omega})^{\alpha}}{\mathbb{G}\eta+\mathbb{V}\eta (i \tilde{\omega})^{\alpha-\beta}+\mathbb{V}\mathbb{G}(i \tilde{\omega})^{\alpha-1}}\,. \label{eq:1}
\end{eqnarray}

Past the gel point, i.e., for $t>t_g$, the solid-like response of the CNC gel is captured by a fractional Kelvin-Voigt model (see upper sketch in Fig.~2 in the main text) composed of the spring-pot ($\mathbb{G}$, $\beta$) characterizing the gel point, in parallel with a purely elastic spring (defined by its elastic modulus $k$) and a second spring-pot ($\mathbb{K}$, $\xi$) with  $0<\xi<\beta<1$. The complex viscoelastic modulus for this model reads \cite{Jaishankar:2013}:
\begin{eqnarray}
     \tilde{G}^*(\tilde{\omega})&=&\tilde{G}'(\tilde{\omega})+i\tilde{G}''(\tilde{\omega})\\
     &=&k+\mathbb{G}(i\tilde{\omega})^{\beta}+\mathbb{K}(i\tilde{\omega})^{\xi}\,. \label{eq:2}
\end{eqnarray}

In order to get more physical insight into these two models, let us examine their high- and low-frequency limits. First, at high rescaled frequencies, since $0<\beta<\alpha<1$ and $0<\xi<\beta<1$, both the fractional Maxwell model and the fractional Kelvin-Voigt model lead to a power-law evolution of the complex viscoelastic modulus with exponent $\beta$, $\tilde{G}^*(\tilde{\omega}) = \mathbb{G}(i\tilde{\omega})^{\beta}$, so that $\tilde{G}'(\tilde{\omega}) \sim \tilde{G}''(\tilde{\omega}) \sim \mathbb{G} \tilde{\omega}^{\beta}$. This corresponds to the power-law rheology observed at the critical gel point, i.e., to the frequency-independent plateau of $\tan \delta(\tilde{\omega})=\beta\pi/2$ at $t=t_g$. 

Second, considering the recovery dynamics in the liquid-like state prior to the gel point, the limit of low rescaled frequencies in the fractional Maxwell model corresponds to short times after flow cessation. There, to leading order in $\tilde{\omega}\rightarrow 0$, the fractional Maxwell model given by Eq.~\eqref{eq:1} yields a linear, viscous scaling for the loss modulus, $\tilde{G}''=\eta \tilde{\omega}$, and a power-law evolution of the elastic modulus, $\tilde{G}' = \frac{\eta^2}{\mathbb{V}}\cos(\alpha\pi/2)\tilde{\omega} ^{2-\alpha}$.

Third, by construction of our rescaled data, long times after the gel point, i.e., late rebuilding dynamics of the solid-like state, correspond to the limit of low rescaled frequencies of the fractional Kelvin-Voigt model. For $\tilde{\omega}\rightarrow 0$, Eq.~\eqref{eq:2} leads to a frequency-independent elastic modulus, $\tilde{G}'=k$, and to a power-law evolution of the viscous modulus, $\tilde{G}'' =\mathbb{K}\sin(\xi\pi/2) \tilde{\omega} ^{\xi}$. 

\subsection*{\large Data analysis and fitting parameters}

\begin{figure*}[!h]
    \centering\includegraphics{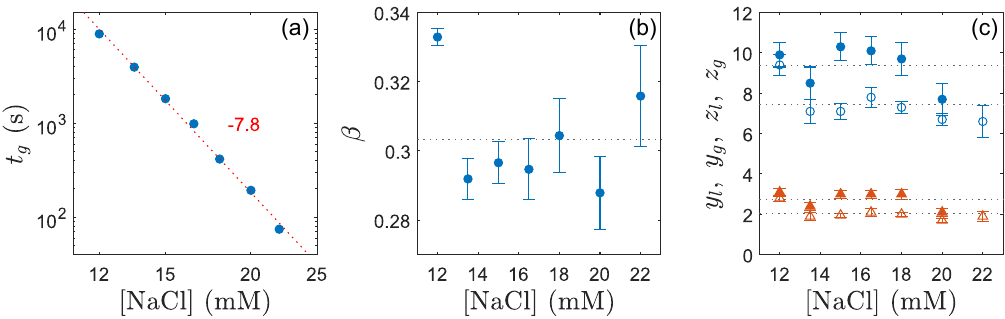}
    \caption{(a) Gelation time $t_g$, (b)~exponent $\beta$ of the power-law frequency dependence of the viscoelastic moduli at the gel point, and (c)~critical exponents $y_l$ ($\bullet$) and $y_g$ ($\circ$) extracted from the time evolution of the frequency shift factor $a(t)$ respectively before and after the gel point, together with the corresponding exponents $z_l$ ($\blacktriangle$) and $z_g$ ($\triangle$) extracted from the moduli shift factor $b(t)$. The gray dashed lines represent the mean values: $\langle\beta\rangle=0.304$, $\langle y_l\rangle=9.37$, $\langle y_g\rangle=7.43$, $\langle z_l\rangle=2.77$, and $\langle z_g\rangle=2.05$. Data from experiments performed on 2~wt~\% CNC suspensions with various salt contents and plotted against the salt concentration $[\text{NaCl}]$.
    The uncertainty on $t_g$ in (a) is of the order of $\pm 5~\%$ (smaller than the symbol size). Error bars in (b,c) show the ranges of variation of the various exponents when $t_g$ is allowed to vary by about 10~\%.
    }
    \label{fig:parametres_2wtpc}
\end{figure*}

We performed time-resolved spectroscopy experiments on 2~wt~\% CNC suspensions with a salt concentration $[\text{NaCl}]$ ranging from 12 to 22~mM. The  data analysis shown in the main text for $[\text{NaCl}]=15$~mM was applied to all samples. The gelation time $t_g$ was first systematically identified as in Fig.~1 in the main text. Figure~\ref{fig:parametres_2wtpc}(a) shows that $t_g$ steeply decreases as a power-law of $[\text{NaCl}]$ with an exponent of about -8, consistently with previous reports on CNC suspensions \cite{Peddireddy:2016,Morlet-Decarnin:2022}. Despite this considerable acceleration of the dynamics, it is possible to distinguish between a liquid-like pre-gel state for $t<t_g$ and a solid-like post-gel state for $t>t_g$ for all salt concentrations within the range under study. Therefore, for every salt concentration, we follow the procedure described in the main text, based on the data-driven automated method recently introduced in Ref.~\cite{Lennon:2023}, to (i)~construct master curves for $\tan\delta$ by rescaling the frequency into $\tilde{\omega}=a(t)\omega$, then (ii)~obtain  master curves for the elastic and viscous moduli by rescaling both $G'$ and $G''$ into $\tilde{G}'=b(t)G'$ and $\tilde{G}''=b(t)G''$. 

\begin{figure*}[!h]
    \centering
    \includegraphics[width=0.75\textwidth]{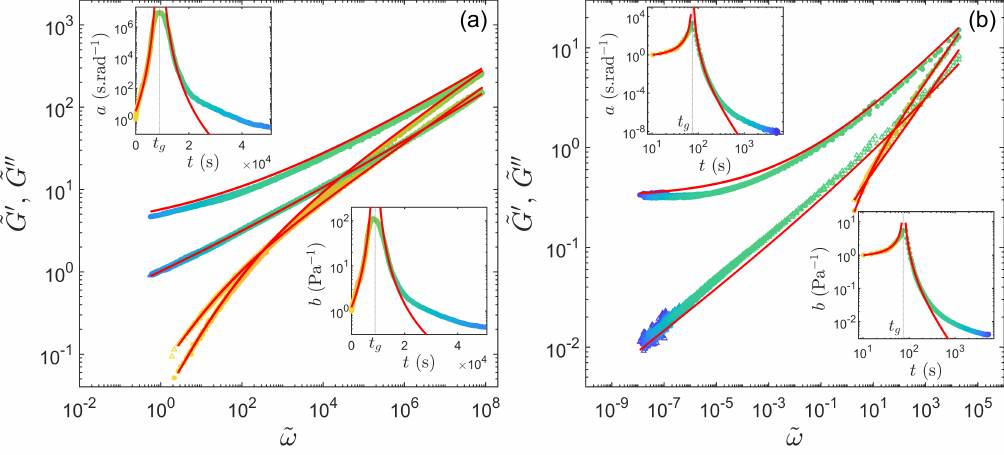}
    \caption{Master curves obtained on 2~wt~\% CNC aqueous dispersions with (a)~$[\text{NaCl}]=12$~mM and (b)~$[\text{NaCl}]=22$~mM, by rescaling the elastic modulus $G'$ ($\bullet$) and the viscous modulus $G''$ ($\triangle$) by the same multiplicative factor $b(t)$, i.e., $\tilde{G}'=b(t)G'$ and $\tilde{G}''=b(t)G''$. The horizontal axis is the rescaled frequency $\tilde \omega=a(t)\omega$ as determined from the construction of master curves for $\tan \delta$ (see procedure in Fig.~1 in the main text). The red lines show the best fits of the data by a fractional Maxwell model for $t<t_g$ [see Eq.~\eqref{eq:1}] and by a fractional Kelvin-Voigt model for $t>t_g$ [see Eq.~\eqref{eq:2}]. Insets: time dependence of the shift factors $a$ (upper insets) and $b$ (lower insets). The initial values are arbitrarily chosen as $a(0)=1$~s.rad$^{-1}$ and $b(0)=1$~Pa$^{-1}$. The red curves show the best power-law fits of the data, $a\sim\epsilon^{-y}$ and $b\sim\epsilon^{-z}$, where $\epsilon=|t-t_g|/t_g$. The exponents are $y_l=9.9$ and $z_l=3.1$ for $t<t_g$ and $y_g=9.4$ and $z_g=2.8$ for $t>t_g$ in (a), and $y_g=6.6$ and $z_g=1.9$ for $t>t_g$ in (b) where the fast gelation prevents one from accurately estimating exponents for $t<t_g$. The vertical dashed lines highlight the gelation times $t_g=8888$~s in (a) and $t_g=75$~s in (b).}
    \label{fig:G_12_22mM}
\end{figure*}

Figure~\ref{fig:G_12_22mM} displays $\tilde{G}'(\tilde{\omega})$ and $\tilde{G}''(\tilde{\omega})$ for the samples with the lowest and highest salt concentrations, respectively 12~mM [Fig.~\ref{fig:G_12_22mM}(a)] and 22~mM [Fig.~\ref{fig:G_12_22mM}(b)]. For all salt concentrations, we found that the above fractional models provide very good fits of the master curves both in the pre-gel state and in the post-gel state (see red lines in Fig.~\ref{fig:G_12_22mM}). Note that since we are dealing with rescaled moduli, the fit parameters that correspond to a viscosity ($\eta$), to quasi-properties ($\mathbb{V}$, $\mathbb{G}$, and $\mathbb{K}$), and to an elastic modulus ($k$) are all dimensionless. In practice, we first fit the viscoelastic spectra very close to $t_g$ by power laws $G'\sim G''\sim \omega^\beta$ to extract the exponent $\beta$. We then fit the master curves [$\tilde{G}'(\tilde{\omega})$,  $\tilde{G}''(\tilde{\omega})$] for $t<t_g$ and $t>t_g$ separately, while imposing the common value of $\beta$ found previously. This yields two values of $\mathbb{G}$ that coincide within 7~\% for all the 2~wt~\% CNC suspensions studied in this work. Thus, in order to lower the number of free parameters and to get a fully consistent modeling of the critical point on both sides of the gel point, we fit again the master curves by imposing the same value of ($\mathbb{G}$, $\beta$) in both the fractional Maxwell and Kelvin-Voigt models, where $\mathbb{G}$ is the average of the two estimations found previously.

Table~\ref{tab:parametres_fit_2wtpc} gathers all the fit parameters for the various salt concentrations. In particular, the parameters $\alpha$, $\beta$, and $\xi$, i.e., the orders of the fractional derivatives involved in the various spring-pots, which control the viscoelastic behavior in the various low- and high-frequency limits as detailed above, seem remarkably independent of the salt content [see also Fig.~\ref{fig:parametres_2wtpc}(b) for $\beta$]. The good agreement with the experimental data demonstrates that the fractional Maxwell model (for $t<t_g$) and the fractional Kelvin-Voigt model (for $t>t_g$) remain valid when varying the salt content. Moreover, we found that, whatever the salt concentration, the shift factors $a(t)$ and $b(t)$ display a critical-like divergence with the time to gelation, i.e., $a(t)~\sim\epsilon^{-y}$ and $b(t)\sim\epsilon^{-z}$, where $\epsilon=|t-t_g|/t_g$. Figure~\ref{fig:parametres_2wtpc}(c) shows the critical exponents $y_i$ and $z_i$ extracted respectively from $a(t)$ and $b(t)$, where $i=l$ ($i=g$ resp.) denotes the pre-gel liquid-like state (post-gel solid-like state resp.). For the largest salt content at $[\text{NaCl}]=22$~mM, the gelation gets too fast [see insets in Fig.~\ref{fig:G_12_22mM}(b)] and exponents cannot be accurately estimated for $t<t_g$. Overall, the exponents $y_i$ and $z_i$ do not follow any systematic trend with $[\text{NaCl}]$: they remain about three times smaller for $b(t)$ than for $a(t)$, with average values $\langle y_i\rangle=8.4$ and $\langle z_i\rangle=2.4$, which is consistent with the hyperscaling relations discussed in the main text. The divergence of the frequency shift factor is thus much more dramatic than that of the moduli shift factor. Moreover, the exponents prior to gelation ($y_l$ and $z_l$) are always about 20~\% larger than the exponents characterizing the post-gel dynamics ($y_g$ and $z_g$), indicating an asymmetrical divergence around the gel point.

\clearpage
\newpage

\begin{table*}[t]
    \centering
\begin{tabularx}{\textwidth}{| >{\centering\arraybackslash}X | >{\centering\arraybackslash}X | >{\centering\arraybackslash}X | >{\centering\arraybackslash}X | >{\centering\arraybackslash}X | >{\centering\arraybackslash}X | >{\centering\arraybackslash}X | >{\centering\arraybackslash}X | >{\centering\arraybackslash}X | }
    \hline
    $[\text{NaCl}]$ (mM) & $\eta$ & $\mathbb{V}$ & $\alpha$ & $\mathbb{G}$ & $\beta$ & $\mathbb{K}$ & $\xi$ & $k$ \rule[-7pt]{0pt}{20pt}\\
    \hline
    12 & 0.17 & 0.13 & 0.69 & 0.54 & 0.33 & 2.38 & 0.20 & 3.00 \\
    13.5 & 1.00 & 0.19 & 0.60 & 0.35 & 0.29 & 0.60 & 0.18 & 0.49 \\
    15 & 0.13 & 0.15 & 0.63 & 0.95 & 0.30 & 5.39 & 0.23 & 17.00 \\
    16.5 & 0.15 & 0.14 & 0.60 & 1.10 & 0.29 & 3.00 & 0.20 & 8.00 \\
    18 & 0.23 & 0.14 & 0.63 & 1.21 & 0.30 & 5.90 & 0.19 & 12.00 \\
    20 & 0.52 & 0.34 & 0.59 & 0.85 & 0.29 & 2.90 & 0.18 & 4.60 \\
    22 & 0.52 & 1.14$\times 10^4$ & 0.70 & 0.40 &  0.32 & 1.10 & 0.20 & 0.32 \\
    \hline
\end{tabularx}
    \caption{Parameters of the fractional Maxwell and Kelvin-Voigt models used to fit the viscoelastic moduli and $\tan \delta$ master curves obtained from the time-evolution of the viscoelastic spectra during the recovery following a strong shear of aqueous suspensions containing 2~wt~\% of CNC and various NaCl concentrations. The typical uncertainty on the fit parameters is 10~\%. The very large value of $\mathbb{V}$ for $[\text{NaCl}]=22$~mM is most probably an outlier due to the restricted amount of data available in the pre-gel state for $t<t_g$ in this fast-gelling sample [see Fig.~\ref{fig:G_12_22mM}(b)].}
    \label{tab:parametres_fit_2wtpc}
\end{table*}

\end{document}